\documentclass[twocolumn,showpacs,preprintnumbers,amsmath,amssymb]{revtex4}

\input{epsf}

\usepackage{amssymb}
\usepackage{amsmath}
\usepackage{epsfig}
\usepackage{amsfonts}
\newsavebox{\astrutbox}
\sbox{\astrutbox}{\rule[-5pt]{0pt}{20pt}}

\begin{document}

\title{Collisional effects on the numerical recurrence in Vlasov-Poisson simulations}

\author{Oreste Pezzi$^1$, Enrico Camporeale$^2$ and Francesco Valentini$^1$}
\affiliation{
$^1$Dipartimento di Fisica and CNISM, Universit\`a della Calabria, 87036 Rende (CS), Italy.\\
$^2$Center for Mathematics and Computer Science (CWI), 1090 GB Amsterdam, Netherlands}
%\pacs{52.20.-j; 52.20.Fs; 52.25.Dg; 52.65.-y; 52.65.Ff}

\begin{abstract}
The initial state recurrence in numerical simulations of the Vlasov-Poisson system is a well-known phenomenon. Here we study the effect on recurrence of artificial
collisions modeled through the Lenard-Bernstein operator [A. Lenard and I. B. Bernstein, Phys. Rev. {\bf 112}, 1456--1459 (1958)]. By decomposing the linear
Vlasov-Poisson system in the Fourier-Hermite space, the recurrence problem is investigated in the linear regime of the damping of a Langmuir wave and of the onset of the
bump-on-tail instability. The analysis is then confirmed and extended to the nonlinear regime through a Eulerian collisional Vlasov-Poisson code. It is found that,
despite being routinely used, an artificial collisionality is not a viable way of preventing recurrence in numerical simulations without compromising the kinetic nature
of the solution. Moreover, it is shown how numerical effects associated to the generation of fine velocity scales, can modify the
physical features of the system evolution even in nonlinear regime. This means that filamentation-like phenomena, usually
associated with low amplitude fluctuations contexts, can play a role even in nonlinear regime.

\end{abstract}

\date{\today}
\maketitle

\section{Introduction}
\label{secintro}
When the Vlasov-Poisson equations are studied by means of Eulerian numerical simulations, one encounters, for low amplitude fluctuations, the problem of the
initial state recurrence. As explained by Cheng et al. \cite{cheng76}, the recurrence phenomenon is intimately related to the presence of a free-streaming term in the
distribution function and to the filamentation problem \cite{zabusky65,cheng76,galeotti05,califano06,cheng13}. Since the mesh-size of the velocity grid is necessarily
finite, the initial state is periodically re-constructed, and thus the electric field exhibits a fake recurrence of the initial state, whose period is $T_{rec} =
2\pi/k\Delta v$, $k$ being the perturbation wavenumber and $\Delta v$ the numerical grid mesh in velocity space. 

In this paper, the effects of collisions on the phenomenon of the numerical recurrence are discussed. Collisions are modeled through the Lenard-Bernstein (LB)
operator, firstly proposed in 1958 by Lenard and Bernstein \cite{lenard58} as a full three-dimensional velocity space collisional operator. The LB operator is a linear
Fokker-Planck collisional operator which belongs, as the Dougherty one \cite{dougherty64,dougherty67}, to the class of ``simplified'' collisional operators and both
collisional terms can be interpreted as advection-diffusion operators in the velocity space. Basically, since both theoretical and numerical approaches of the
Landau collisional integral \cite{landau36,rosenbluth57,akhiezer86} - the natural collisional operator for plasmas - are very complicated, effects of collisions are
usually modeled through simplified collisional operators as the two mentioned above. Recently, by comparing the effects of the Landau and Dougherty operators in a
homogeneous plasma, Pezzi et al. \cite{pezzi15a} have shown that the two operators can be successfully compared once time is opportunely scaled by a constant factor. This
represents a quite good and computationally affordable way to perform self-consistent collisional simulations in the realistic three-dimensional velocity space
\cite{pezzi15b}.

However, when collisions act on longitudinal electrostatic waves and the system dynamics occurs preferentially in a unique direction, one can describe collisional
effects in a reduced one-dimensional velocity space by adopting one-dimensional collisional operators. This approach has already been adopted in several works
\cite{zk63,su68,oneil68,anderson07a,anderson07b,pezzi13,pezzi14a,pezzi14b,camporeale15}. In the same spirit, here we focus on the one-dimensional LB operator. Our
analysis complements previous results discussed in earlier works (see Refs. \cite{ng99,ng04,siminos11,black13}): as these authors pointed out, for the linear Landau
damping problem an opportune collisionality can prevent the onset of the numerical recurrence and restore the correct collisionless damping as expected within the Landau
theory \cite{landau46}. 

Moreover, an important result in the study of the LB operator has been established by Ng et al. \cite{ng99}: while the collisionless Vlasov-Poisson system supports a
continuous spectrum of neutral singular eigenmodes (the so-called Case-Van Kampen modes \cite{vankampen55,case59}), the introduction of the LB operator modifies the
spectrum into a set of proper, discrete eigenmodes. 
The Landau damping phenomenon, which in the collisionless case appears due to the phase-mixing of the
continuous spectrum \cite{landau46,krall86}, is recovered as a discrete eigenmode, in the limit of vanishing collisionality.
However, as we will show, when the Vlasov-Poisson-LB system is discretized in the velocity space (and hence bound to a finite
resolution), the Landau root is recovered as a discrete eigenvalue only for a given value of collisionality.

Interestingly, the same effect on the spectrum induced by LB collisions has been discussed in Ref. \cite{siminos11} in the context of {\em spectral deformation}. This
is a technique introduced for the Vlasov-Poisson system in \cite{crawford89}, where a non-unitary transformation is applied to the linear operator, in such a way that its
eigenvalues with nonzero real part remain unchanged, while the continuum of neutral modes is damped. In analogy to the case of LB operator, the Landau damping is
recovered as a true eigenmode. Therefore, we suggest that the LB operator might be interpreted as a spectral deformation to the collisionless Vlasov-Poisson system.
However, the precise identification of the transformation which is equivalent to the LB operator is left for future work.

The aim of our analysis is to understand if recursive effects or any other numerical effect associated to limited velocity resolution of Eulerian calculations can be
successfully removed by making use of a collisional operator, without increasing the number of gridpoints in the velocity domain (and without altering the physical
features of the system evolution). In case of positive response, this would be extremely useful especially for multi-dimensional simulations, where the velocity
resolution is limited for computational reasons. We show that, in general, the collision frequency $\nu$ which is suitable for preventing recurrence in the linear regime
is a function of the perturbation wavenumber: as the wavenumber increases a stronger collisionality is necessary to avoid the onset of the numerical recurrence. Moreover,
by focusing on the nonlinear Landau damping and in particular on the formation of a Bernstein-Greene-Kruskal (BGK) nonlinear wave \cite{bgk57,oneil65}, we show that 
i) the collisionless case is also slightly affected by recurrence and ii) collisional effects become important when the dynamics evolve to the nonlinear stage.
Therefore, it seems impossible to use the LB operator to avoid the numerical recurrence and, simultaneously, preserve the phase space structures developed as in the
collisionless case. Of course, in the case of higher velocity resolution, for which the recurrence time is significantly
larger than the characteristic time of the physical process of interest (Landau damping, onset of instabilities, generation of
nonlinear BGK structures and so on), the use of a collisional operator opportunely tailored to eliminate numerical recurrence
does not affect the reliability of the physical results for times smaller than the recurrence time. However, let us remark that
this case is not the one of interest in our analysis in which we intentionally choose to have recurrence in the initial stage of
the simulations, which typically cannot afford a very fine resolution in velocity space (especially in multi-dimensions).
Finally, by exploring the recurrence effect on the bump-on-tail instability \cite{krall86}, we show that the recurrence affects both the linear
exponential growth and the nonlinear saturation of the instability by producing a fake growth in the electric field and that, as in the nonlinear Landau damping case,
collisional effects are not able to prevent the initial state recurrence without significantly altering the nonlinear structures.

In summary, the purpose of this paper is twofold. First, we show how recursive effect and filamentation, which are usually described in the context of low amplitude
fluctuations, can also be problematic in nonlinear phenomena, such as the saturation regime of the bump-on-tail instability. Second, we discuss a useful diagnostic, in
terms of expansion of the velocity space into Hermite functions, that allows to better appreciate the effect of an artificial collisional operator in phase space.

Let us summarize the content of the paper. In Sec. \ref{secmath} the theoretical background of the problem is given and the numerical strategies adopted to approach the
solution are explained. Then, in Sec. \ref{secLD}, the recurrence effects on the Landau damping phenomenon are described in both linear and nonlinear regimes by
transforming the Vlasov-Poisson system into Hermite-Fourier coordinates and by means of Eulerian simulations. Moreover, we investigate how collisional effects prevent the
recurrence problem but, at the same time, smooth out the nonlinear plasma dynamics features as the system evolves to the nonlinear regime. Then, in Sec. \ref{secBUMP}
we analyze the initial state recurrence problem and the collisional effects for the case of the bump-on-tail instability. Finally, in Sec. \ref{secconcl} we conclude by
summarizing the shown results. 

\section{Theoretical background and numerical models}
\label{secmath}
Here we consider a quasi-neutral and unmagnetized plasma composed by kinetic electrons and immobile background ions. We assume that only electrostatic interactions occur
between particles, therefore the Maxwell system reduces to the Poisson equation. Furthermore, since electron-ion and ion-ion collision frequencies are much smaller than
the electron-electron one, we take into account only electron-electron collisions \cite{akhiezer86}. As introduced above, electron-electron collisions are modeled through
the LB collisional operator \cite{lenard58}.

The normalized collisional Vlasov-Poisson (VP) equations - where collisions are modeled through the LB collisional operator - in the 1D--1V (one dimension in physical
space and one dimension in velocity space) phase space configuration reads:
\begin{eqnarray}
& & \frac{\partial f}{\partial t}+v \frac{\partial f}{\partial x}+\frac{\partial\phi}{\partial x}\frac{\partial f}{\partial v}=\left. \frac{\partial f}{\partial
t}\right|_{coll} \label{vlaseq}\\
&-&\frac{\partial^2 \phi}{\partial x^2} =  1 -\int f\ dv \ ; \label{poiseq}
\end{eqnarray}
where $f=f(x,v)$ is the electron distribution function, $\phi(x)$ is the electrostatic potential, defined as $E=-d\phi/dx$ ($E$ is the electric field) and $\left. \partial
f/\partial t \right|_{coll}$ is the LB collisional operator. Due to their inertia, the protons are considered as a motionless neutralizing background of constant density
$n_0=1$. In previous equations, time is scaled to the inverse electron plasma frequency $\omega_{pe}$, velocities to the initial electron thermal speed $v_{th,e}$;
consequently, lengths are normalized by the electron Debye length $\lambda_{De}=v_{th,e}/\omega_{pe}$ and the electric field by $\omega_{pe} m v_{th,e}/e$ ($m$ and $e$
being the electron mass and charge, respectively). For the sake of simplicity, from now on, all quantities will be scaled using the characteristic parameters listed
above.

The scaled Lenard-Bernstein \cite{lenard58} collision operator is:
\begin{equation}
\left. \frac{\partial f}{\partial t}\right|_{coll}  = \nu \frac{\partial }{\partial v} \left[ \frac{\partial f}{\partial v} + v  f \right] \;
\label{lbeq}
\end{equation}
being $\nu$ the constant collisional frequency. The LB operator preserves global mass. Moreover, if the distribution function has null average speed $V=0$ and unitary
temperature $T=1$, being $V=1/n \int dv f v$, $n=\int dv f$ and $T=1/n \int dv (v-V)^2 f $ respectively plasma mean velocity, density and temperature, it conserves also
momentum and energy. 

In the following we analyze the equations system Eqs. \eqref{vlaseq}--\eqref{poiseq} coupled to Eq. \eqref{lbeq}. For the sake of simplicity, we refer to this system
compactly as Eqs. \eqref{vlaseq}--\eqref{poiseq}. Two different analyses have been performed on Eqs. \eqref{vlaseq}--\eqref{poiseq} and are briefly explained in the
following two subsections.

\subsection{Fourier-Hermite decomposition (Linear analysis)}
\label{secHF}
A very convenient way of studying the properties of the LB operator in the linear regime is by employing an expansion of the linearized distribution function into a
Fourier-Hermite basis. Here, we use the so-called asymmetrically weighted Hermite functions \cite{schumer98,camporeale06,camporeale15}:
\begin{eqnarray}
\Psi_n(\xi) &=& (\pi2^nn!)^{-1/2}H_n(\xi)e^{-\xi^2}\label{psi_1}\\
\Psi^n(\xi) &=& (2^nn!)^{-1/2}H_n(\xi),\label{psi_2}
\end{eqnarray}
where $H_n$ is the n-th Hermite polynomial, defined as
\begin{equation}
 H_n(\xi) = (-1)^n e^{\xi^2} \frac{d^n}{d\xi^n}\left(e^{-\xi^2}\right),
\end{equation}
and $\xi=v/\sqrt{2}$. 
The basis in Eqs. \eqref{psi_1}--\eqref{psi_2} has the following properties:
\begin{eqnarray}
\int^\infty_{-\infty} \Psi_n(\xi)\Psi^m(\xi) d\xi = \delta_{n,m},\\
v\Psi_n(\xi) = \sqrt{n+1}\Psi_{n+1}(\xi)+\sqrt{n}\Psi_{n-1},\\
\frac{d\Psi_n(\xi)}{dv} = -\sqrt{(n+1)}\Psi_{n+1}(\xi),
\end{eqnarray}
$\delta_{n,m}$ being the Kronecker delta.
Eqs. \eqref{vlaseq}--\eqref{poiseq} are linearized around an homogeneous equilibrium that, when expanded in Hermite functions, reads
$f_0(v)=\sum_{n=0}C_n^{eq}\Psi_n(\xi)$. Note that, for a Maxwellian equilibrium with zero mean velocity, all coefficients $C_n^{eq}$ are null for $n>0$. The perturbed
distribution function $f_1(x,v)=f(x,v)-f_0(v)$ is expanded as:
\begin{equation}
 f_1(x,v) = \sum_{n,j} C_{n,j} \Psi_n\left(\frac{v}{\sqrt{2}}\right)e^{\mathrm{i}k_j x},\label{expansion}
\end{equation}
with $k_j=2\pi j/L$, and $L$ the domain length. By using the orthogonality of the Fourier-Hermite basis, one obtains, for each $k_j$ mode:

\begin{equation}
 \begin{split}
  & \frac{dC_{n,j}}{d t}  + \mathrm{i}k_j\Biggl(\sqrt{n+1}C_{n+1,j} +  \Biggr. \\
  & \Biggl. \sqrt{n} C_{n-1,j} + \frac{\sqrt{2n}}{k_j^2}C_{0,j}C_{n-1}^{eq}\Biggr) +n\nu C_{n,j}  =0
 \end{split}
\label{linear_eq}
\end{equation}

Note that $\Psi_n(\xi)$ is an eigenfunction of the LB operator of Eq. \eqref{lbeq}, with eigenvalue $n\nu$, and thus the use of the rescaling factor in the argument of 
the basis in Eqs. \eqref{psi_1}--\eqref{psi_2} allows to obtain a rather compact formulation (compare, for instance, with the formulation in \cite{black13}). In
particular, the linear equation (\ref{linear_eq}) can be written in matrix form as:
\begin{equation}
 \frac{d \overrightarrow{C_j}}{dt} =  \mathbf{A}_j \overrightarrow{C_j},\label{matrix_eq}
\end{equation}
where $\overrightarrow{C_j}$ is the vector defined as $(C_{0,j},C_{1,j},C_{2,j},\ldots)^T$, and the matrix $\mathbf{A}_j$ is defined as
\small
\begin{equation}
\mathbf{A}_j = -\mathrm{i}k_j
 \scriptscriptstyle{\begin{pmatrix}
  0 & 1 & 0 &  \\
  1+\sqrt{2}C_0^{eq}/k_j^2 & \nu/\mathrm{i}k_j &  \sqrt{2} & 0  \\
  2C_1^{eq}/k_j^2 & \sqrt{2}  & 2\nu/\mathrm{i}k_j   & \sqrt{3} & 0 &   \\
  \sqrt{6}C_2^{eq}/k_j^2 & 0 & \sqrt{3}  & 3\nu/\mathrm{i}k_j   & \sqrt{4} & \ddots \\
  \ddots & \ddots & \ddots & \ddots & \ddots & \ddots
 \end{pmatrix}}
\end{equation}
\normalsize
The collisionality $\nu$ affects only the diagonal entries of the matrix. Once again, this is due to the fact that the Hermite basis is an eigenfunction of the LB
operator. Of course, when numerically solving the linear problem in Eq. \eqref{matrix_eq}, one has to truncate the matrix $\mathbf{A}$, that is, one has to choose the
maximum number $N_H$ of Hermite modes in the expansion of Eq. \eqref{expansion}, by setting $C_{n,j}=0$ for any $n> N_H$ (other closures have been investigated, see, e.g.
\cite{gibelli06,dasilva13}). This corresponds to defining the resolution in velocity space. It is precisely the inability to capture increasingly finer scales in velocity
space that gives rise to the phenomenon of recurrence in the numerical solutions of Vlasov equation. This becomes particularly clear by looking at the recurrence effect
within the framework of the Hermite basis expansion in velocity.

\begin{figure}
\epsfxsize=8.5cm \centerline{\epsffile{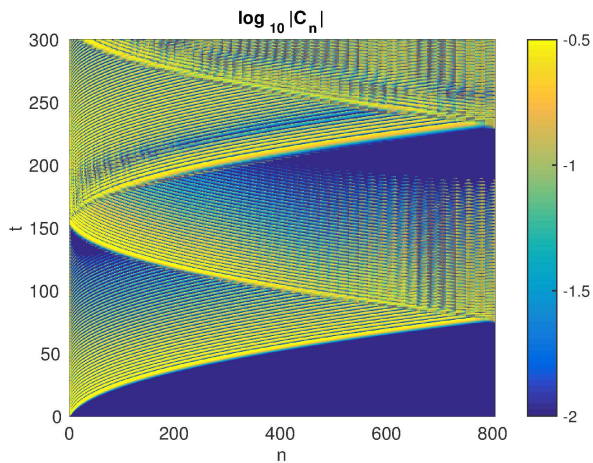}}   % FIGURE N.1
\caption{(Color online) Temporal evolution of the Hermite coefficients $|C_n|$ (in logarithm scale) as a function of the Hermite mode $n$ and the time $t$ for the
collisionless $\nu=0$ case.}
\label{fig1}
\end{figure}

\subsection{Eulerian Vlasov code (nonlinear analysis)}
\label{secVPcode}
The other approach consists in the numerical solution of Eqs. \eqref{vlaseq}--\eqref{poiseq} through a Eulerian code based on a finite difference scheme for the
approximation of spatial and velocity derivatives of $f$ over the grid-points. Time evolution of the distribution function is approximated through the splitting scheme
first introduced by Filbet et al. \cite{filbet02} [see Refs. \cite{pezzi13,pezzi14a} for details about the numerical algorithm], which is a generalization of the
well-known splitting scheme discussed by \cite{cheng76}. We impose periodic boundary conditions in physical space and $f$ is set equal to zero for $|v|> v_{max}$, where
$v_{max}=6v_{th,e}$. Phase space is discretized with $N_x$ grid points in the physical domain and $N_v$ points in the velocity domain. Finally, the time step $\Delta t$
has been chosen in such a way to respect the Courant-Friedrichs-Levy condition \cite{peyret83} for the numerical stability of time explicit finite difference schemes. 

\begin{figure}[!b]
\epsfxsize=8.5cm \centerline{\epsffile{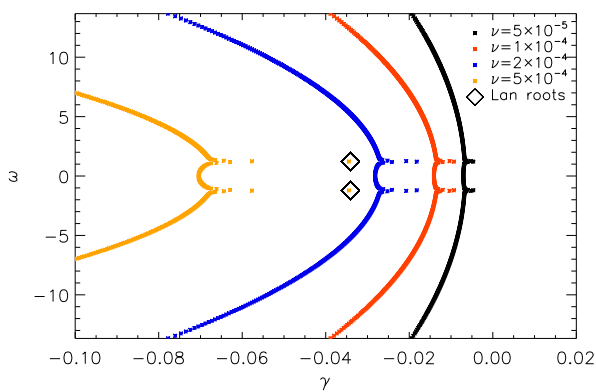}}   % FIGURE N.2
\caption{(Color online) Spectrum of the matrix $\mathbf{A}$ for four increasing values of collisionality: $\nu=5\times10^{-5},1\times10^{-4},2\times10^{-4},
5\times10^{-4}$ respectively in black, red, blue and gold dots. The black squares represent the Landau roots.}
\label{fig2}
\end{figure}

The plasma is initially in an equilibrium state and we perturb the system through an oscillating density perturbation which produces, through the Poisson equation, a
perturbative electric field of amplitude $\delta E$.

\section{Landau damping}
\label{secLD}
In the present section, recurrence effects and collisional effects on this phenomenon are described for the the case of the Landau damping of a Langmuir wave.

First, we study a collisionless ($\nu=0$) linear Landau damping case, for the wavenumber $k=k_1=2\pi/L=0.35$ (being $L=18$), by means of the Fourier-Hermite
decomposition with $N_H=800$. The system is initially perturbed through a spatially sinusoidal electric field perturbation, which translates, in the Fourier-Hermite
space, to initialize the vector $\overrightarrow{C_j}$ as $(1,0,0,\ldots)^T$ (the electric field is proportional to $C_{0}$). 

Figure \ref{fig1} shows the temporal evolution of the absolute value of the Hermite coefficients $|C_n|$ in logarithm scale. Since the filamentation in velocity space
naturally produces small velocity scales, Hermite coefficients of increasingly higher modes are excited. When the largest mode gets excited, the truncation of the series
acts as a reflecting boundary (around time $T\sim 75$), and the perturbation travels back towards lower modes. Around time $T\sim 150$, the electric field damping is
abruptly interrupted and a value close to the initial value is restored. Let us note that, although the electric field will not be affected until the recurrence time
$T\sim 150$, the distribution function is spuriously altered from time $T\sim 75$, that is when the perturbation reflects on the boundary.

\begin{figure}[!b]
\epsfxsize=8.5cm \centerline{\epsffile{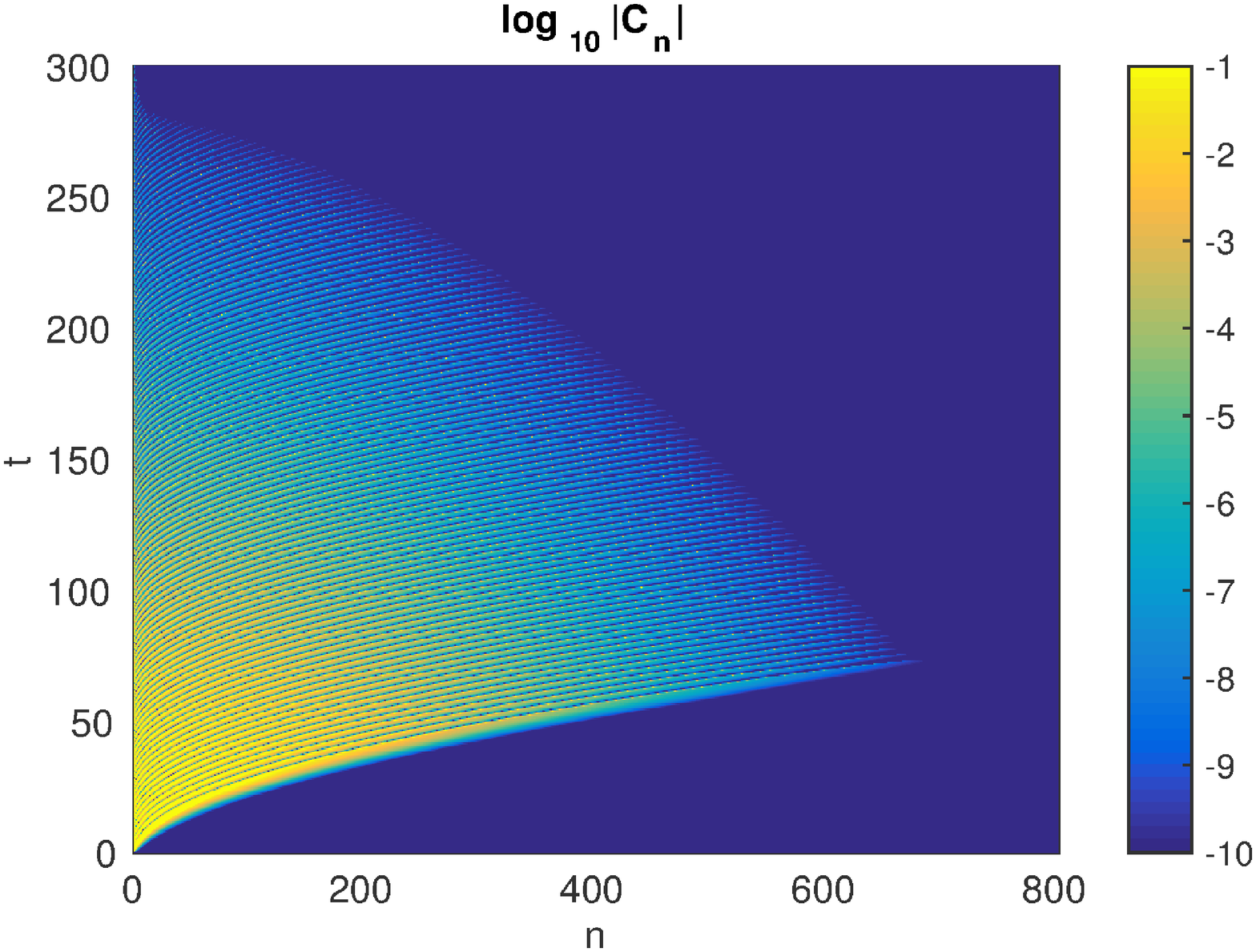}}   % FIGURE N.3
\caption{(Color online) Temporal evolution of the Hermite coefficients $|C_n|$ as a function of the Hermite mode $n$ and the time $t$ for the collisional
$\nu=5\times10^{-4}$ case.}
\label{fig3}
\end{figure}

As we mentioned earlier, the effect of a non-null collisionality in the Vlasov-Poisson linear operator is to modify the spectrum of eigenvalues. Landau damping is not
anymore due to the phase-mixing of a continuous set of neutral mode. Moreover, for a large enough value of $\nu$, it appears as the least-damped eigenvalue of the system.
This is shown in Figure \ref{fig2}, where, for the same value of $k=k_1=0.35$, we show the spectrum of the matrix $\mathbf{A}$ for four increasing values of
collisionality: $\nu=5\times10^{-5},1\times10^{-4},2\times10^{-4}, 5\times10^{-4}$ (respectively in black, red, blue and gold dots). The damping rate $\gamma$ and the
wave propagation frequency $\omega$ are respectively shown on the horizontal and vertical axes of Fig. \ref{fig2}. The values corresponding to the theoretical Langmuir
roots ($\gamma=\gamma_L=-3.37\times10^{-2}$ and $\omega=\pm 1.22$), obtained through the numerical evaluation of the Landau dispersion function roots, are shown as
black squares. 
We emphasize that the spectrum of the matrix $\mathbf{A}$ differs from the spectrum of the infinite-dimensional
Vlasov-Poisson-LB operator. In fact, while for the latter the Landau root is a discrete eigenvalue in the limit $\nu\rightarrow
0$, Figure 2 clearly shows that, in the presence of a finite velocity resolution, a small collisionality acts to distort the
discrete representation of the Case-Van Kampen continuum. In other words, a sufficiently large collisionality value (depending on
the velocity resolution) is needed in order to recover the Landau root as a discrete mode. Indeed, it is clear that, for
$\nu=5\times10^{-4}$ (gold points), the spectrum exhibits two eigenvalues overlapping with the proper Landau roots value and,
therefore, the proper Landau damping is restored. 

In order to clarify the behavior of the coefficients $|C_n|$ in the case where the collisionality restores the proper Landau damping (i.e. $\nu=5\times10^{-4}$), we show
in Fig. \ref{fig3} the temporal evolution of the Hermite coefficients $|C_n|$. Clearly the reflecting effect discussed for Fig. \ref{fig1} has now completely vanished and
the electric field damping does not show any recurrence. Since the collisional operator damps the high Hermite modes or, in other words, since collisional effects stop
the production of small velocity scales, the velocity filamentation is not correctly captured. 

\begin{figure}
\epsfxsize=8.5cm \centerline{\epsffile{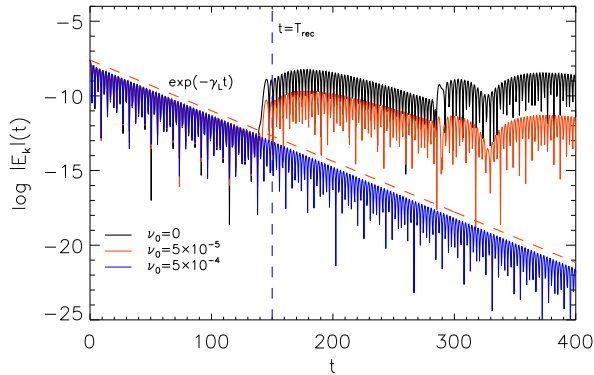}}   % FIGURE N.4
\caption{(Color online) Temporal evolution of $\log |E_k|(t)$ with $k=k_1$. The black, red and blue lines indicate respectively  $\nu=0$,  $\nu=5\times10^{-5}$ and 
$\nu=5\times10^{-4}$. The red and blue dashed lines show respectively the theoretical damping with Landau damping $\gamma_L$ and the instant time $t=T_{rec}$.}
\label{fig4}
\end{figure}

In order to complete our analysis, we numerically solve Eqs. \eqref{vlaseq}--\eqref{poiseq} through the finite-difference numerical code presented earlier, for different
values of the collisional frequency $\nu$. We set the initial sinusoidal density perturbation such that the perturbation electric field amplitude is $\delta E= 10^{-3}$.
The phase space is discretized with $N_x=64$ and $N_v=101$ points. Let us remark that, with the parameters choice just described, the recurrence time is
$T_{rec}=2\pi/k \Delta v \simeq 150$.

The time evolution of the logarithm of the absolute value of the first Fourier component $k=k_1$ of the electric field $\log |E_{k}|(t)$ is shown in Fig. \ref{fig4}. The
black, red and blue lines correspond respectively to the collisionless case ($\nu = 0$), intermediate collisional case ($\nu=5\times10^{-5}$) and stronger collisional
case ($\nu= 5\times 10^{-4}$). The last case is the case in which the Landau damping root is recovered in the spectrum shown in Fig. \ref{fig2}, thanks
to the effect of collisions. The red and blue dashed lines in Fig. \ref{fig4} indicates the theoretical Landau damping rate $\gamma_L=-3.37\times10^{-2}$ and the
recurrence time $t=T_{rec}\simeq150$ respectively. 

For the three cases, the electric field spectral component evolution is approximately the same for $t<T_{rec}$ and the electric field is damped at the proper Landau
damping rate $\gamma_L$. Then, around $t=T_{rec}\simeq150$, the collisionless and the intermediate collisional cases (black and red solid lines of Fig. \ref{fig4})
present a fake ``jump'' in the signal due to the initial state recurrence problem. On the other hand, in the stronger collisional case $\nu=5\times10^{-4}$ (blue solid
line of Fig. \ref{fig4}), the recurrence effect disappears and the unphysical ``jump'' is completely suppressed by collisional effects. It is worth to note that, in this
case, the recurrence does not occur neither at times multiples of the recurrence period. 

\begin{figure}[!b]
\epsfxsize=8.5cm \centerline{\epsffile{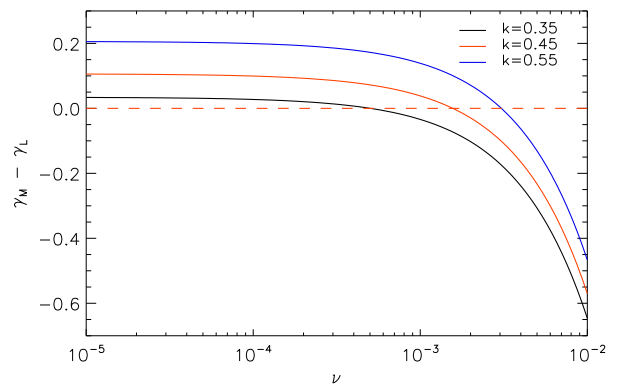}}  % FIGURE N.5
\caption{(Color online) The black, red and blue lines show the difference between the damping rate $\gamma_M$ of the least damped mode and the damping rate $\gamma_{L}$
of the Landau root, as a function of the collisional rate $\nu$, for three different values of $k=0.35, 0.45, 0.55$ respectively.}
\label{fig5}
\end{figure}

Based on the results presented above, the inclusion of a weakly collisional operator to prevent the numerical recurrence effect might look convenient; however, the
consequences of including collisionality into the Vlasov-Poisson system must be investigated with care. 

\begin{figure*}
\epsfxsize=17cm \centerline{\epsffile{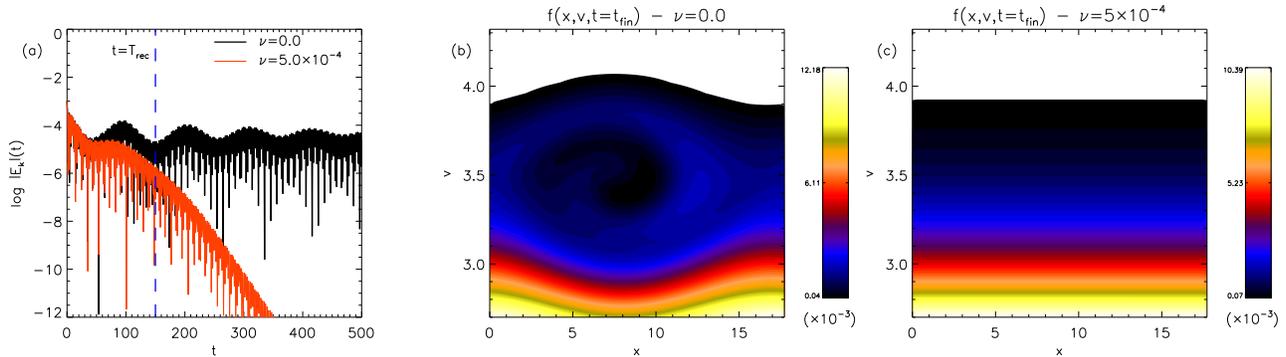}}   % FIGURE N.6
\caption{(Color online) (a) Temporal evolution of $\log |E_k|(t)$ with $k=k_1$ for the collisionless case (black line) and the collisional $\nu=5\times10^{-4}$ case (red
line). The blue dashed vertical line indicates the recurrence period $T_{rec}$. The distribution function around the phase speed $v=v_\phi$ at the final time instant
$f(x,v,t=t_{fin})$ is shown in panels (b)--(c) for the collisionless (b) and collisional (c) case.}
\label{fig6}
\end{figure*}

Figure \ref{fig5} shows the difference between the damping rate $\gamma_M$ of the least damped mode and the damping rate $\gamma_{L}$ of the Landau root, as a
function of the collisional rate $\nu$, for three different values of $k=0.35, 0.45, 0.55$, (black, red and blue line,
respectively). As explained in Figure \ref{fig2}, for $\nu\rightarrow 0$, and fixed velocity resolution, the Case-Van Kampen
spectrum \cite{vankampen55,case59} is recovered (see Fig. \ref{fig2}), and $\gamma_M\rightarrow0$. The intersection between the
red dashed and the solid lines indicates the value of collisionality that is required to recover the correct Landau damping as a discrete eigenmode. Moreover, bearing in
mind that both $\gamma_M$ and $\gamma_{L}$ are negative quantities, values above the red-dashed line in the figure indicate that the collisional rate is not large enough
to recover the Landau damping as the least damped eigenvalue, while values below the red-dashed line indicate over-damping with respect to the Landau damping. Figure
\ref{fig5} clearly indicates that there is not a single value of collisionality that would allow to recover the correct Landau damping for a spectrum of several
wavenumbers. Since larger wavenumbers are subject to stronger damping, they would require a larger collisional rate 

Moreover, if the initial field amplitude is increased in order to explore the nonlinear evolution of the Landau damping, the collisionality, which was able of preventing
recurrence in the linear simulation, becomes strong enough to smooth the nonlinear physical features of the Landau damping. In order to clarify this point, we perform a
simulation with the same parameters of the linear one explained above (see Fig. \ref{fig4}) and we increase $\delta E=10^{-1}$. Figure \ref{fig6} (a) shows the time
evolution of $\log |E_{k}|(t)$ for $k=k_1$ for the collisionless case (black solid line) and for the collisional case $\nu=5\times 10^{-4}$ (red solid line). { The
blue dashed line in Fig. \ref{fig6} indicates the recurrence period $T_{rec}=2\pi/k \Delta v \simeq 150$. We remark that this specific value of collisional frequency is
the one which prevents recurrence effects in the linear case, still preserving the correct value of Landau damping}. 

{ It is clear that, in the non-linear collisionless case, the Landau damping is arrested by nonlinear effects (particle trapping) and, as a consequence, the electric
field starts oscillating around a nearly constant saturation level. On the other hand, in the collisional case, the physical scenario changes drastically and the electric
field amplitude displays evident collisional damping. }

In phase space, nonlinear effects manifest as the generation of a vortical trapping population, moving with velocity close to the wave phase speed ($v_\phi\simeq3.50$).
This is shown in Figs \ref{fig6} (b)--(c) where the contour plots of the distribution function $f(x,v)$ at time $t=400$ for the collisionless case (b) and for the
collisional case (c) are reported. It is clear from the comparison of panels (b) and (c) of Fig. \ref{fig6} that collisions prevent the generation of the phase-space
trapping population, since they work to smooth out any deformation of the particle distribution function and to drive the system toward thermal equilibrium. In other
words, as soon as kinetic effects produce distortions (and, consequently, sharp velocity gradients) of the particle distribution, collisional effects become more intense
to keep the  velocity distribution close to a Maxwellian. Therefore, it is quite clear that collisional effects are not able to prevent the recurrence problem without
destroying the plasma dynamics characteristics. 

\begin{figure}[!b]
\epsfxsize=9cm \centerline{\epsffile{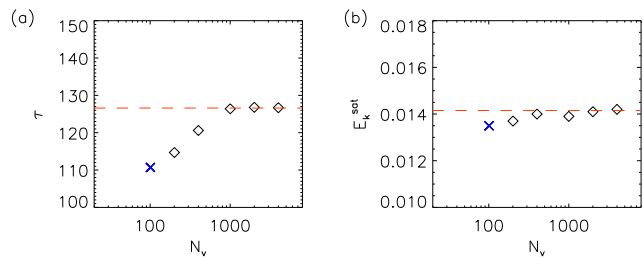}}   % FIGURE N.7
\caption{(Color online) The oscillation period of the electric field envelope $\tau$ (a) and the saturation electric field $E_{k,sat}$ (b) as a function of $N_v$. The
blue crosses indicate the $N_v$ case case depicted in Fig. \ref{fig6}.}
\label{fig7}
\end{figure}

In order to understand whether changing the resolution in velocity space \cite{galeotti05,califano06} affects the physical features of the system, we performed additional
simulations in collisionless regime, increasing the number of gridpoints in the velocity domain: $N_v=101, 201, 401, 1001, 2001, 4001$; $N_v=101$ [indicated with blue
crosses in Figs. \ref{fig7} (a--b)] corresponds to the case depicted in Fig. \ref{fig6}.

We computed the following quantities as ``proxies'' of numerical accuracy:
\begin{itemize}
 \item The oscillation period $T_{osc}$ of the wave, evaluated in the time interval $t\ge T_{rec}$);
 \item The time $t_{max}$ where the electric field envelope reaches its first maximum [$\simeq100$ in Fig. \ref{fig6}(a)];
 \item The oscillation period $\tau$ of the electric field envelope, defined as the average of the difference between two consecutive maximum points in the $\log
|E_{k}|(t)$ evolution; 
 \item The saturation electric field $E_{k,sat}$ at which the electric field spectral power saturates.
 \end{itemize}
The quantities $T_{osc}$ and $t_{max}$ (not shown here) do not depend on $N_v$, the relative variations between the two extremes cases ($N_v=101$ and $N_v=4001$) being
always smaller than the $1\%$. On the other hand, in Fig. \ref{fig7} we report the dependence of $E_{k,sat}$ (a) and $\tau$ (b) on $N_v$. Clearly, these two quantities
approach a saturation value (red-dashed line) as $N_v$ increases. The relative variations between the values obtained with $N_v=101$ and the corresponding saturation
values (red dashed lines) are about the $4\%$ for $E_{k,sat}$ and $10\%$ for $\tau$. We conclude that even in the nonlinear case shown in Fig. \ref{fig6} the limited
resolution in the velocity domain slightly affects the physical evolution of the system. However, as discussed above, adding a collisional operator to eliminate these
unphysical effects produces drastic changes in the kinetic aspects of the dynamics with respect to the collisionless case.

\section{Bump-on-tail instability}
\label{secBUMP}
In the current section the recurrence effects on the bump-on-tail instability are described by performing a similar analysis to that performed in Sec. \ref{secLD}. The
initial distribution function is the following:
\begin{eqnarray}\nonumber
& &f_0(v)= \frac{n_0}{(2\pi T_0)^{1/2}} \exp\left(-\frac{v^2}{2T_0}\right) + \frac{n_b}{(2\pi T_b)^{1/2}}\times \\
& &\Bigg[\exp \left(-\frac{(v - V_b)^2}{2T_b}\right) +\exp \left(-\frac{(v_x + V_b)^2}{2T_b}\right)\Bigg]
\label{bumpfd}
\end{eqnarray}
The core density and temperature are respectively $n_0=0.98$ and $T_0=1$, while the bump density, mean velocity and temperature are $n_b=0.01$, $V_b=4$ and $T_b=0.4$
respectively. Is it clear that $f_0(v)$ represents a Maxwellian distribution function to which two bumps are superimposed at both positive and negative side of the
velocity domain. Moreover the velocity symmetry in the velocity shape of $f_0(v)$ guarantees an initial null current. In Hermite space, the parity of $f_0(v)$ translates
to having $C_n^{eq}=0$ for all odd $n$.

\begin{figure}[!b]
\epsfxsize=8.5cm \centerline{\epsffile{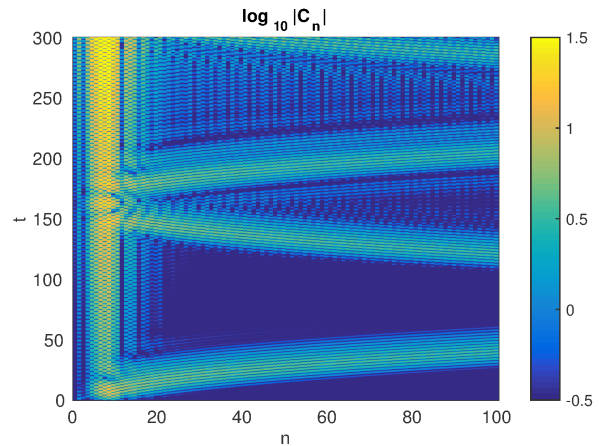}}   % FIGURE N.8
\caption{(Color online) Temporal evolution of the Hermite coefficients $|C_n|$ as a function of the Hermite mode $n$ and the time $t$ for the collisionless $\nu=0$ case.}
\label{fig8}
\end{figure}

First of all, as performed in Sec. \ref{secLD}, we study the collisionless ($\nu=0$) linear evolution of the bump-on-tail instability onset for $k=k_1=2\pi/L=0.25$
(being the plasma length $L=25$) by perturbing initially the system through a spatially sinusoidal electric field perturbation. Here the Hermite mode numbers is
$N_H=400$. Figure \ref{fig8} shows the temporal evolution of the absolute value of the Hermite coefficients $|C_n|$. Only the first $100$ modes are shown, to better
appreciate the recurrence on the low order modes. As in Fig. \ref{fig1} for the Landau damping, the filamentation creates small velocity scales and, due to the truncation
of the Hermite series - which corresponds, in the Eulerian code, to the presence of a finite velocity grid size - the boundary reflects back the perturbation towards
lower modes. The main difference with respect to the Landau damping case is that now there is an eigenmode whose amplitude grows exponentially in time. The eigenmode has
a certain structure in Hermite space, and is localized between modes 5 and 10. Once the filamentation bounces back because of the truncation of the series, the unstable
eigenmode is perturbed, around time $T\sim 150$. Therefore, in the bump-on-tail case, the recurrence is much more evident as a fake perturbation acting on the unstable
eigenmode, rather than on the electric field. In fact, as we show in the following, the recurrence of the electric field is more modest than for the Landau damping case.

In order to clarify how the recurrence acts on the instability onset, we perform some Eulerian simulations where the phase space is discretized with $N_x=128$ point while
$N_v$ is variable in order to change the recurrence period: $N_v=101$ ($T_{rec}\simeq200$), $N_v=201$ ($T_{rec}\simeq400$) and $N_v=1001$ ($T_{rec}\simeq2000$). We
perturb the system through a sinusoidal density perturbation whose wavenumber is $k=k_1=0.25$. The density perturbation amplitude is $\delta n= 2.51 \times10^{-6}$ which
corresponds to a perturbed electric field of amplitude $\delta E=10^{-5}$. By evaluating the dispersion function roots of the Vlasov equation we can calculate, for the
specific wavenumber, the linear growth rate of the instability $\gamma_I^{th}=9.20\times10^{-3}$ and the wave phase speed $v_{\phi}=3.90$.

\begin{figure}
\epsfxsize=8.5cm \centerline{\epsffile{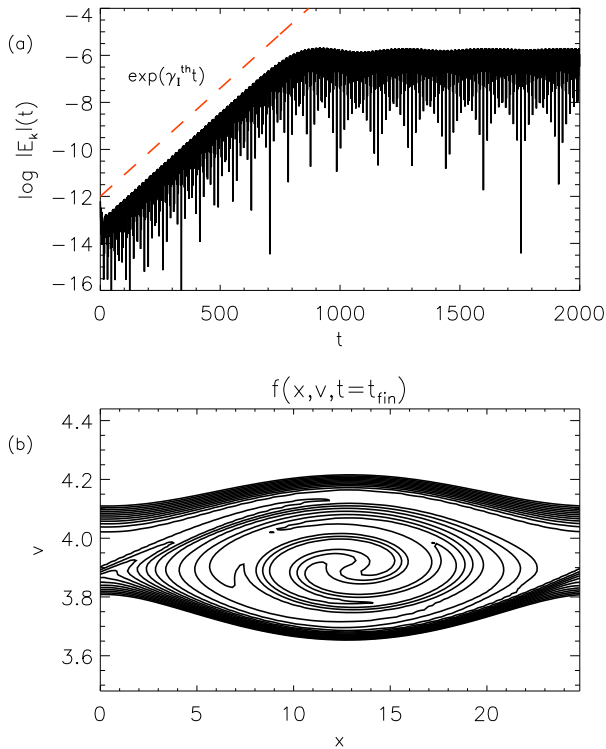}}   % FIGURE N.9
\caption{(Color online) (a) Temporal evolution of $\log |E_k|(t)$ with $k=k_1$ for the collisionless recurrence-free ($N_v=1001$) case. The red dashed line represents the
theoretical growth expectation $\exp(\gamma_I^{th} t)$. (b) Contour plot of the distribution function around the phase space $v=v_\phi$ at the final time instant
$f(x,v,t=t_{fin})$.}
\label{fig9}
\end{figure}

Figures \ref{fig9} (a)--(b) show respectively the temporal evolution of $\log |E_k|(t)$ with $k=k_1$ and the phase space contour plot at the final time of the simulation
$t=t_{fin}$ for the high resolution case ($N_v=1001$). Clearly the instability is not affected by the recurrence and, in the linear stage, the field amplitude grows up
exponentially in accordance with the theoretical prediction [red dashed line in Fig. \ref{fig9} (a)]. As nonlinear effects become important, the field saturates at a
constant value and in the phase space, a BGK-like structure \cite{oneil65,bgk57} is formed [see Fig. \ref{fig9}(b)]. The phase space structure is well-localized around
the phase speed $v=v_\phi$ and its width is quite in accordance with the theoretical prediction.

\begin{figure*}
\epsfxsize=17cm \centerline{\epsffile{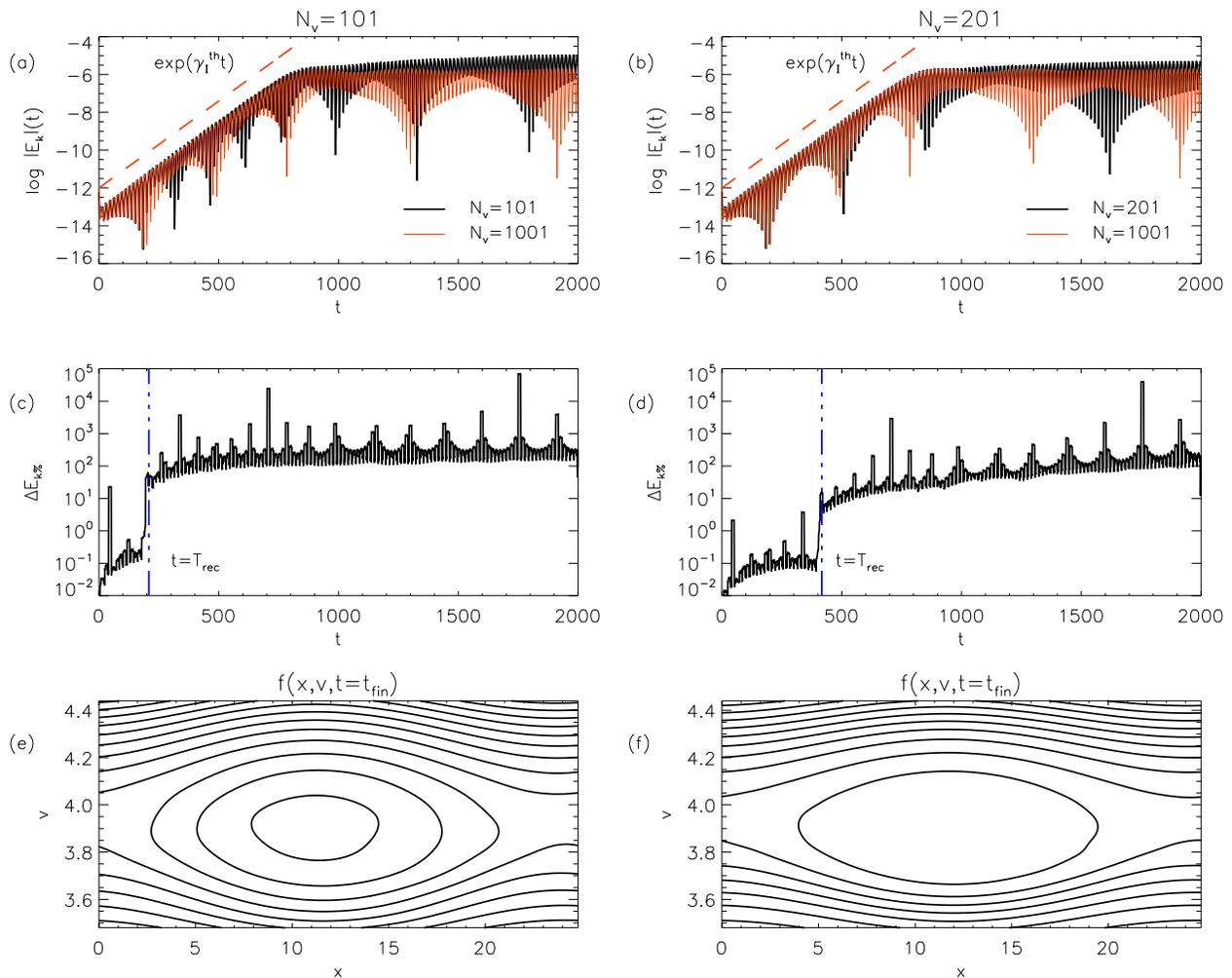}}   % FIGURE N.10
\caption{(Color online) Recurrence effects on the bump-on-tail instability for the $N_v=101$ (left column) and $N_v=201$ (right column) simulations. The top panels
(a)--(b) show the temporal evolution of $\log |E_k|(t)$ with $k=k_1$ for the low-resolution case (black line) and for the recurrence-free case (red solid line), while the
red dashed line indicates the theoretical growth expectation $\exp(\gamma_I^{th} t)$. The central panels (c)--(d) display the quantity $\Delta E_{k\%}$ (black line) and
the recurrence period $t=T_{rec}$ (blue dashed line). Finally the bottom panels (e)--(f) visualize the distribution function contour plot around the phase space
$v=v_\phi$ at the final time instant $f(x,v,t=t_{fin})$.}
\label{fig10}
\end{figure*}

In contrast with the case just shown, when the velocity resolution decreases, recursive effects occur. Panels of Figs. \ref{fig10} show the results of two simulations
with resolution $N_v=101$ (left column) and $N_v=201$ (right column). For each column, the top panel [Figs. \ref{fig10} (a)--(b)] describes the temporal evolution of
$\log |E_k|(t)$, while the center panel [Figs. \ref{fig10} (c)--(d)] displays the quantity $\Delta E_{k\%}$, defined as the relative difference (expressed in percentage)
between $|E_k|(t)$ at a given resolution and $|E_k|(t)$ for the collisionless recurrence-free case. Finally, the bottom contour plot [Figs. \ref{fig10} (e)--(f)] exhibits
the distribution function $f(x,v,t=t_{fin})$ at the final time and around the phase speed $v=v_\phi$. Let us remark that, in order to better visualize the phase space
structures in Fig. \ref{fig10} (e)--(f), we performed an interpolation of the distribution function over a more resolved grid without altering the physical features of
the phase space structure itself.

\begin{figure}
\epsfxsize=8.5cm \centerline{\epsffile{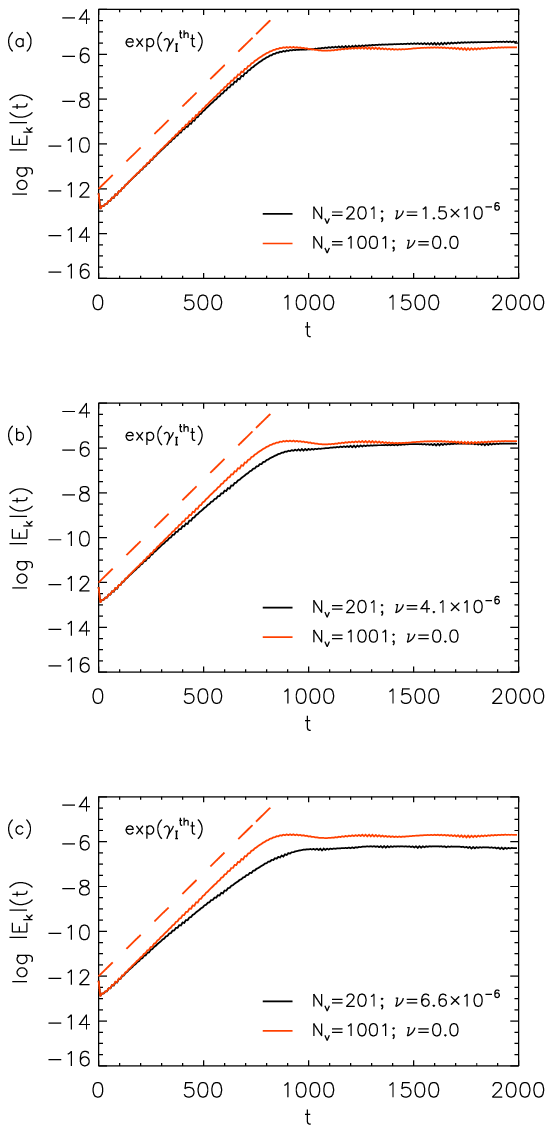}}   % FIGURE N.11
\caption{(Color online) Temporal evolution (black line) of $\log |E_k|(t)$ with $k=k_1$ for the case $N_v=201$ and with collisional frequency $\nu=1.5\times10^{-6}$ (a),
$\nu=4.1\times10^{-6}$ (b) and $\nu=6.6\times10^{-6}$ (c) respectively. In each panel the red solid line shows the evolution of $\log |E_k|(t)$ for the collisionless
recurrence-free ($N_v=1001$) case while the red dashed line displays the theoretical linear instability growth.}
\label{fig11}
\end{figure}

It is clear that the recurrence also manifests in the instability onset. By focusing on the linear stage of the instability growth, the electric field amplitude seems to
exponentially increase at a rate in accordance with the theoretical expectations, represented with red dashed lines in Figs. \ref{fig10} (a)--(b). Moreover, as introduced
above, in contrast with the Landau damping case, the recurrence effect does not strongly manifest as a fake jump around the recurrence time $t=T_{rec}$. However, by
analyzing the temporal evolution of $\Delta E_{k\%}$ [see Figs. \ref{fig10} (c)--(d)], a abrupt increase of $\Delta E_{k\%}$ is observed around the recurrence period,
shown in Figs. \ref{fig10} (c)--(d) with blue dashed lines. This discontinuity is due to recursive effects and it means that, after the recurrence period, the electric
field evolution in the case with a lower resolution strongly departs from the recurrence-free case ($\Delta E_{k\%} \simeq 100\%$). Thus, although recursive effects
cannot be appreciated in the linear stage of the instability growth by looking directly at Figs. \ref{fig10} (a)--(b) (the scale is logarithmic and a variation about the
$100\%$ cannot be easily highlighted), the field evolution is actually disturbed by recurrence. 

Furthermore, recurrence phenomena affect the nonlinear evolution of the instability. Effectively, by focusing on Fig. \ref{fig10} (a)--(b), in the case without
recurrence the electric field power opportunely saturates at a constant value (red line) while, on the other hand in the cases with recurrence the electric field
does not saturate and it continues to slowly increase. Finally, by focusing on the distribution function at the final time instant $t=t_{fin}$ [see Figs. \ref{fig10}
(e)--(f)], in both cases a phase space structure is produced around the correct phase speed. By comparing these phase space structures with the hole created in the
recurrence-free case [Fig. \ref{fig9} (b)], some differences clearly reveal. First, phase space structures obtained in the cases with recurrence are less resolved
compared to the one of the recurrence-free case and this is obviously related to the different velocity grid size: effectively, since the velocity grid size is smaller in
the recurrence-free case, finer scales are naturally created compared to the cases at lower resolution. Moreover, the vortex width seems to be slightly wider in the
$N_v=101$ case [Fig. \ref{fig10} (e)] compared to both the collisionless recurrence-free case [Fig. \ref{fig9} (b)] and to the $N_v=201$ case [Fig. \ref{fig10} (f)]. In
other words, since the electric field does not saturate in presence of recursive effects, the phase space structure tends to increase its width.

The effects of the initial state recurrence on the bump-on-tail instability represents a novel and quite unexpected feature in the analysis of the recursive phenomena.
Both linear and nonlinear stages of the instability are affected by recurrence: the electric field evolution departs from the evolution in the case without
recurrence ($N_v=1001$) around $t=T_{rec}$. Furthermore the nonlinear saturation, which is properly retained in the case at high resolution, is interrupted by recurrence
as the velocity grid size gets larger. Moreover, due to the absence of the electric field saturation, the distribution function shows a vortex properly centered around
the right phase speed but whose width tends to be bigger compared to the case without recurrence. Finally, although initial state recurrence phenomena are often related
to linear physical problems, here we have found some new and interesting recurrence effect features which occur in the nonlinear regime. 

In order to explore if a collisionality described by the LB operator could represent a good way to prevent numerical recurrence in the case of the bump-on-tail
instability, we focus on the $N_v=201$ resolution case ad we perform several collisional simulations by changing the collisional frequency $\nu$.

Figs. \ref{fig11} (a)--(c) display, through black lines, the temporal evolution of $\log |E_k|(t)$ with $k=k_1$ for the cases: $\nu=1.5\times 10^{-6}$ (a),
$\nu=4.1\times10^{-6}$ (b) and $\nu=6.6\times 10^{-6}$ (c). In each panel of Fig. \ref{fig11} red solid lines indicate the evolution in the collisionless case without
recurrence [the same shown in Fig. \ref{fig9} (a) and in Figs. \ref{fig10} (a)--(b)] while the red dashed line shows the theoretical expectation for the instability
growth
curve $\exp (\gamma_I^{th} t)$, being $\gamma_I^{th}=9.2\times10^{-3}$.

As expected, collisions inhibit the instability and tend to restore thermal equilibrium. However in the case $\nu=1.5\times 10^{-6}$ [see Fig. \ref{fig11} (a)],
collisions weakly affect the electric field evolution which, as in the collisionless case, do not saturate and overtake the recurrence-free case evolution [red
line in Fig. \ref{fig11} (a)]. 

As collisional frequency increases, the electric field evolution tends to be dissipated. In the intermediate case $\nu=4.1\times10^{-6}$ [see Fig. \ref{fig11} (b)], the
electric field reaches, at the end of the simulation, almost the same power of the collisionless case without recurrence; however its evolution departs from the reference
red curve around $t\simeq600$, where the recurrence-free case [red line in Fig. \ref{fig11} (b)] presents a stronger power level than the collisional $N_v=201$ case
[black line in Fig. \ref{fig11} (b)]. On the other hand, in the case $\nu=6.6\times 10^{-6}$ [see Fig. \ref{fig11} (c)], a significant difference between the two
evolutions appears at even smaller time instants and collisions clearly affect the linear instability regime. In particular, the linear growth rate in the collisional
$N_v=201$ case [black line in Fig. \ref{fig11} (c)] is significantly smaller than the collisionless $N_v=1001$ case [red line in Fig. \ref{fig11} (c)]. Moreover, as in
the collisionless recurrence-free case, at the final stages of the simulation the electric field spectral power exhibits an almost flat behavior at a lower power value
compared to the collisionless recurrence-free case.

\begin{figure}[!b]
\epsfxsize=8cm \centerline{\epsffile{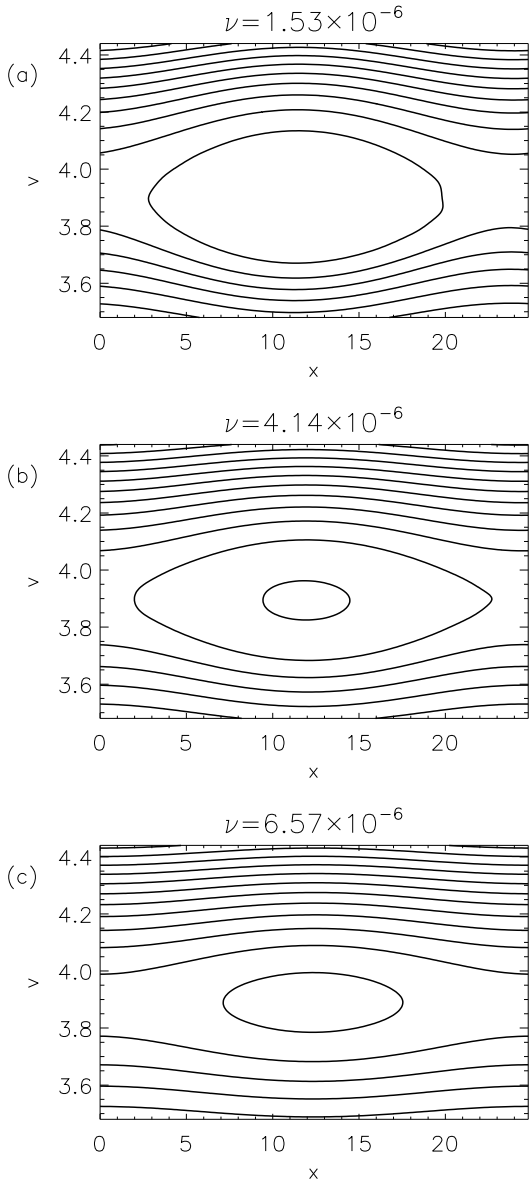}}   % FIGURE N.12
\caption{(Color online)The distribution function contour plots around the phase space $v=v_\phi$ at the final time instant $f(x,v,t=t_{fin})$ for the case $N_v=201$ and
with collisional frequency $\nu=1.5\times10^{-6}$ (a), $\nu=4.1\times10^{-6}$ (b) and $\nu=6.6\times10^{-6}$ (c).}
\label{fig12}
\end{figure}

In order to point out how phase space is affected by collisions, Figs. \ref{fig12} (a)--(c) show the contour plots of the distribution function $f(x,v,t=t_{fin})$ at
the final time instant $t=t_{fin}$ and zoomed around the phase speed $v=v_\phi$ for the cases: $\nu=1.5\times 10^{-6}$ (a), $\nu=4.1\times10^{-6}$ (b) and $\nu=6.6\times
10^{-6}$ (c). As in Fig. \ref{fig10} (e)--(f), even in Fig. \ref{fig12} (a)--(c) we performed an interpolation of the distribution function over a more resolved
grid. In all the three cases shown in Fig. \ref{fig12} (a)--(c) a phase space structure is observed around the wave phase speed and its width reduces as collisional
frequency increases. Clearly as collisions become stronger, phase space structures are smoothed out and present a smaller size.

To conclude this section, we highlight that, as collisional frequency gets bigger, the instability is affected by collisions more intensely. Moreover, since collisions
tend to restore the equilibrium, they have been active since the initial stage of the simulation (the initial distribution function is out of equilibrium).
Furthermore they remain active until the equilibrium is recovered and incessantly work to smooth out all the wave features (electric field signal and phase space
structures). Therefore, at longer times (not shown here), the phase space structures shown in Figs. \ref{fig12} get smaller and disappear, while the electric field signal
shown in Figs. \ref{fig11} is dissipated by collisional effects. We conclude that, as in the nonlinear Landau damping case, an artificial collisionality is not able to
prevent the initial state recurrence in the bump-on-tail instability onset. In particular we found two different scenarios: collisions are so weak that recurrence is
still active or, on the other hand, they affect both recurrence effects and physical evolution of the system by deeply smoothing the electric field and the phase space
structure. 

\section{Conclusions}
\label{secconcl}
In this paper we analyzed in detail the problem of the initial state recurrence in a weakly collisional plasma, where electron-electron collisions have been modeled
through the Lenard-Bernstein collisional operator \cite{lenard58}. We focused on two study cases: the Landau damping of a Langmuir wave and the bump-on-tail instability
onset. For both cases, the analysis in the linear regime has been performed through the decomposition of the linear Vlasov-Poisson system into the Fourier-Hermite space.
In particular, the expansion of the distribution function in terms of Hermite functions separates naturally different velocity scales and it allows to better describe
recursive effects and appreciate the role of the collisional operator in phase space. Moreover, the analysis has been extended to the nonlinear regime through a 1D--1V
Eulerian collisional Vlasov-Poisson code, already tested and used in previous works (see Refs. \cite{pezzi13,pezzi14a}). 

Recently some authors (see Refs. \cite{black13} and references therein) pointed out that an opportune collisionality can prevent the onset of recursive effects and
restore the correct Landau damping. This indication suggested us to investigate whether the inclusion of an artificial collisionality could be used to prevent recurrence
in numerical simulations without the loss of physical details due to collisional effects. However, we have shown that the collisional frequency $\nu$ which is suitable for
preventing numerical recurrence in the linear regime depends on the perturbation wavenumber; furthermore, collisional effects become important when the system evolves to
the nonlinear regime and, for the same value of collisionality which prevents recursive effects in the linear stage, any nonlinear wave is strongly dissipated by
collisional effects.

Finally, we pointed out that numerical effects associated to the generation of fine velocity scales can modify the physical features of the system evolution even in
nonlinear regime. This has been shown by focusing on the nonlinear Landau damping phenomenon and on the bump-on-tail instability both in linear and nonlinear regime. Our
results indicate that filamentation-like and recursive effects, often associated with evolution in linear regime, can also be important in the nonlinear case. We also
conclude that the addition of a collisional operator, with the aim of preventing the recurrence of the initial state and other numerical effects related to limited
resolution in the velocity domain, significantly changes the evolution of nonlinear waves and the corresponding phase space portrait.

\section*{Acknowledgments}
E.C and F.V. acknowledge the participation to the ISSI International Team 292 ``Kinetic Turbulence and Heating in the Solar Wind''. This work has been supported by the 
Agenzia Spaziale Italiana under agreement (o contract)  ASI-INAF 2015-039-R.O “Missione M4 di ESA: Partecipazione Italiana alla fase di assessment della missione THOR“.

\end{document}